# The Simulation, Fabrication Technology and Characteristic Research of Micro-Pressure Sensor with Isosceles Trapezoidal Beam-Membrane[*]

Jing Wu, Xiaofeng Zhao*, Yibo Liu and Dianzhong Wen

*The Key Laboratory of Electronics Engineering, College of Hei longjiang Province, Heilongjiang University,*

*Harbin 150080, China*

*\*zhaoxiaofeng@hlju.edu.cn*

A micro-pressure sensor with an isosceles trapezoidal beam-membrane (ITBM) is proposed in this paper, consisting of a square silicon membrane, four isosceles trapezoidal beams and four piezoresistors. To investigate how the elastic silicon membrane affects pressure sensitive characteristics, a simulation models based on ANSYS 15.0 software were used to analyze the effect of structural dimension on characteristics of pressure sensor. According to that, the chips of micro-pressure sensors were fabricated by micro-electro-mechanical system (MEMS) technology on a silicon wafer with <100> orientation. The experimental results show that the proposed sensor achieves a better sensitivity of 9.64 mV/kPa and an excellent linearity of 0.09%F.S. in the range of 0~3.0 kPa at room temperature and a supply voltage of 5.0 V, with a super temperature coefficient of sensitivity (TCS) about - 2180 ppm/℃ from - 40 ℃ to 85 ℃ and low pressure measurement less than 3.0 kPa.

## 1. Introduction

Nowadays, silicon piezoresistive pressure sensor has been one of the most reported and used micro-electro-mechanical system (MEMS) devices to achieve high accuracy, with a wide range of applications for flexible pressure sensor,[1] medical,[2] automobile industries,[3] and other fields.[4] In addition, there is more and more particular attention paid for researchers in piezoresistive pressure sensor due to its super properties of lower power consumption, higher integration level and the ability for small point/space limited measurement.[5] Moreover, it is common knowledge that the fabrication of the micro-pressure sensor mainly includes two methods to improve its characteristics, i.e. discoveries of new materials and designs of new structures.[6] However, the micro-pressure sensors with new structures are more popular due to its outstanding stress concentration effect on silicon diaphragm comparing with the other ones using new materials. For example, in 1993, Sandmaier *et al.* proposed a square-diaphragm pressure sensor with arectangular center boss to measure low-pressure(0~10 kPa) confirmed by experimental measurement.[7] Thereafter, Kinnell *et al.* proposed a 30 mbar (3.0 kPa) full-scale pressure sensor by introducing a hollow stiffening structure for producing stress concentration effect to silicon diaphragm using an electro-chemical etch stop process, achieving a sensitivity of 20 mV/V and a linearity less than 0.40%F.S..[8] In 2012, Tian *et al.* designed a piezoresistive pressure sensor with a cross beam-membrane (CBM) structure in operating range of 0~5.0 kPa, owning a sensitivity of 7.081 mV/kPa and a nonlinearity of 0.09%F.S..[9] After that, Xu *et al.* reported a novel bossed diaphragm pressure sensor combined with a peninsula-island structure, realizing a sensitivity of 0.066 mV/Pa·V and a non-linearity of 0.33%F.S in the range of 0~500 Pa.[10] Meanwhile, Guan *et al.* introduced a shuriken-structured diaphragm to piezoresistive pressure sensor for monitoring the intraocular pressure (IOP) and the intracranial pressure (ICP), obtaining a sensitivity of 4.72 mV/kPa·V and a linearity of 0.18 %F.S. in the pressure range of 0~3.0 kPa.[11] Recently, Li *et al.* proposed a piezoresistive pressure sensor, consisting of four-grooved membrane combined with rood beam to measure low pressure (0~6.7 kPa), realizing a sensitivity of 30.9 mV/Psi·V (4.612 mV/kPa·V) and a nonlinearity error of 0.21%F.S..[12] Base on the above analysis, using the pressure sensors with new structures can realize the micro-pressure measurement. Thus, optimizing the membrane structure makes it possibility to contribute to the improving of micro-pressure sensor performance by utilizing stress concentration effect.

---

[*]Corresponding author.

To further study the effect of elastic element on the performance of micro-pressure sensors, a sensor with an isosceles trapezoidal beam-membrane (ITBM) structure was proposed in this study. According to the piezoresistive effect and the theory of elastic mechanics, the operation principle of resulted micro-pressure sensor was analyzed. In order to investigate the effects of three types of elastic silicon membranes on the sensitivity of pressure sensor, a simulation models of sensor established based on ANSYS 15.0 software were utilized to make comparative analysis. According to that, the isosceles trapezoidal beam-membrane structural dimension was optimized. On the basis of that, a micro-pressure sensor with optimizing structure size was fabricated on a n-type silicon wafer with <100> orientation by micro-electro-mechanical system (MEMS) technology and its relative characteristics were researched in detail.

## 2. Basic Structure and Operation Principle

### 2.1. *Basic Structure*

The micro-pressure sensor with an ITBM structure is given in Figure 1, where the ITBM is composed by a square silicon membrane, four isosceles trapezoidal beams(ITBs) and four piezoresistors. As seen in Figure 1, the sensor makes up of an elastic element and a sensitive element, where the elastic element consists of a square silicon membrane and four ITBs (Beam 1, Beam 2, Beam 3 and Beam 4). And the sensitive element is a Wheatstone bridge with four piezoresistors ($R_1$, $R_2$, $R_3$ and $R_4$) connected by aluminum electrode at the root of four ITBs, in which $R_1$ and $R_3$ are along <011> orientation, and $R_2$ and $R_4$ are along <0$\bar{1}$1> orientation. In addition, one end of $R_1$ and $R_2$ is connected to $R_4$ and $R_3$ as the output voltage $V_{out2}$ and $V_{out1}$ ends, respectively, and the other ends of $R_1$ and $R_2$ is connected to the supply power $V_{DD}$ as well as the other ends of $R_3$ and $R_4$ are connected to the GND, respectively. $L$ and $H$ are the length and the thickness of the chip, respectively. $l$ is the length of silicon membrane, $h$ is the thickness of elastic element. The lengths of topline and baseline for the ITBs are $d$ and $d_1$, the height and the thickness of ITBs are $l_1$ and $h_1$, respectively.

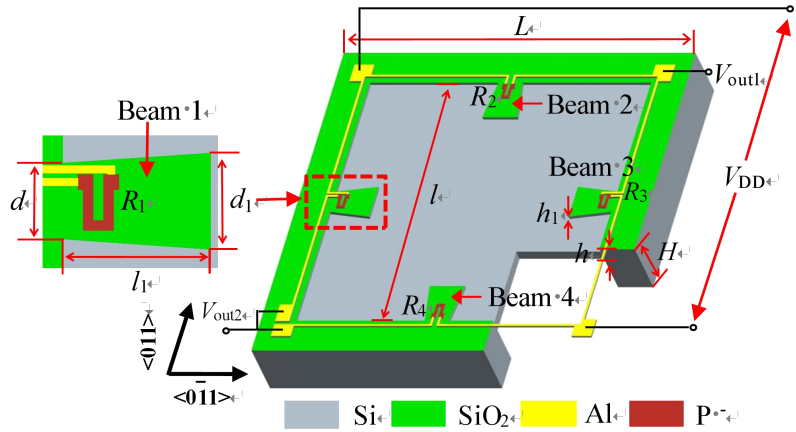

Figure 1. The basic structure of micro-pressure sensor with the ITBM.

## 2.2. *Operation Principle*

Figure 2 (a) and (b) show the operating principle schematic diagram of the proposed sensor. When exerting an external pressure $P$ perpendicular to the surface of the chip, the deformations of the square silicon membrane and four ITBs will cause the resistance changes of the four piezoresistors based on piezoresistive effects. Under the supply voltage $V_{DD}$, the output voltage $V_{out}$ of the Wheatstone bridge is given as following:[13]

$$V_{out} = V_{out2} - V_{out1}$$
$$= \frac{(R_1+\Delta R_1)(R_3+\Delta R_3) - (R_2-\Delta R_2)(R_4-\Delta R_4)}{(R_1+\Delta R_1 + R_4-\Delta R_4)(R_2-\Delta R_2 + R_3+\Delta R_3)} V_{DD} \quad (1)$$

where $R_1$, $R_2$, $R_3$ and $R_4$ are the resistances of four piezoresistors, $\Delta R_1$, $\Delta R_2$, $\Delta R_3$ and $\Delta R_4$ are the variations of $R_1$, $R_2$, $R_3$ and $R_4$, respectively.

In ideal conditions, $R_1=R_2=R_3=R_4=R$, and $\Delta R_1=-\Delta R_2=\Delta R_3=-\Delta R_4$. Under no external pressure as shown in Figure 2 (a), $V_{out}$ of the Wheatstone bridge can be simplified as the equation (2):[14]

$$V_{\text{out}} = V_{\text{out2}} - V_{\text{out1}} = 0 \tag{2}$$

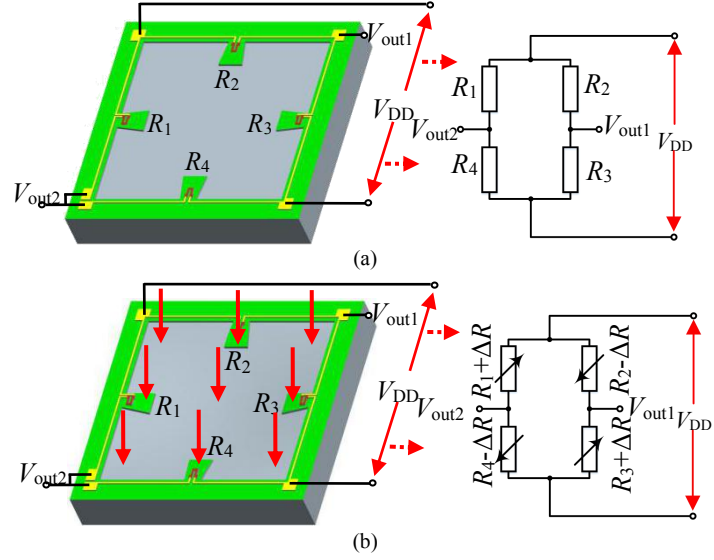

Figure 2. The operation principle of the proposed pressure sensor: (a) under external pressure; (b) under no external pressure.

As shown in Figure 2 (b), when the external pressure $P \neq 0$ kPa, $V_{\text{out}}$ of the Wheatstone bridge is:[15]

$$V_{\text{out}} = V_{\text{out2}} - V_{\text{out1}} = \frac{\Delta R}{R} V_{\text{DD}} \tag{3}$$

According to piezoresistive effect, the relative change of resistances can be expressed as:[16]

$$\frac{\Delta R}{R} = \pi_{\text{ll}} \sigma_{\text{ll}} + \pi_{\perp} \sigma_{\perp} \tag{4}$$

where $\pi_{\text{ll}}$ and $\pi_{\perp}$ are the longitudinal and the transverse piezoresistive coefficients, $\sigma_{\text{ll}}$ and $\sigma_{\perp}$ are the longitudinal and the transverse stresses, respectively.

According to the calculation equations of longitudinal and transverse piezoresistive coefficients in arbitrary orientations, the longitudinal and the transverse piezoresistive coefficients of p-type silicon along <011> orientation on a (100) n-type silicon wafer are:[17]

$$\pi_{\text{ll}} \approx \frac{1}{2} \pi_{44}, \quad \pi_{\perp} \approx -\frac{1}{2} \pi_{44} \tag{5}$$

where $\pi_{44}$ is shear piezoresistive coefficient of silicon. So the $V_{out}$ can be obtained from Eq. (3), (4)and (5):[18]

$$V_{out} = \frac{1}{2}\pi_{44}(\sigma_{\parallel}-\sigma_{\perp})V_{DD} \tag{6}$$

The sensitivity is an important parameter to estimate the performance of a sensor, defined as the relative change of output voltage per unit applied pressure. The sensitivity ($S$) of the sensor can be expressed as:[5]

$$S = \frac{V_{outF} - V_{out0}}{P_F - P_0} = \frac{V_{F.S.}}{P_F - P_0} \tag{7}$$

where $P_F$ and $P_0$ are the maximum and minimum working pressures, respectively; $V_{outF}$ and $V_{out0}$ are the output voltages under $P_F$ and $P_0$, respectively. $V_{F.S.}$ is the full-scale output voltage.

### 3. Simulation and Fabrication

#### 3.1. *Simulation analysis*

##### 3.1.1. *Effects of elastic silicon membrane on pressure sensor characteristics*

To study how the structures affect the characteristics of the pressure sensor, i.e. the square silicon membrane (structure Ⅰ), the beam-membrane (structure Ⅱ) and the isosceles trapezoidal beam-membrane (proposed structure Ⅲ), the input-output characteristics corresponding to the three structures were investigated. On the basis of Mechanical APDL 15.0 module in the ANSYS 15.0 software, the pressure sensor models corresponding to the above mentioned structures were established, with the relative simulation parameters in Table 1. When $V_{DD}$=5.0 V, the relationship curves between the output voltages and the applied pressures of pressure sensors with the three structures are shown in Figure 3, both with the linearly increasing output voltages with the increasing pressure in 0~3.0 kPa. It can be found that the sensor with structure Ⅲ realizes a high sensitivity compared with the conventional structures Ⅰ and Ⅱ.

Table 1. The relative parameters of the three simulation models

| Structure type | Structure Ⅰ | Structure Ⅱ | Structure Ⅲ |
|---|---|---|---|
| $L$ (μm) | 5000 | 5000 | 5000 |
| $l$ (μm) | 4000 | 4000 | 4000 |
| $h$ (μm) | 42 | 42 | 42 |
| $l_1$ (μm) | 580 | 580 | 580 |
| $h_1$ (μm) | — | 7 | 7 |
| $d$ (μm) | — | 340 | 280 |
| $d_1$ (μm) | — | 340 | 340 |

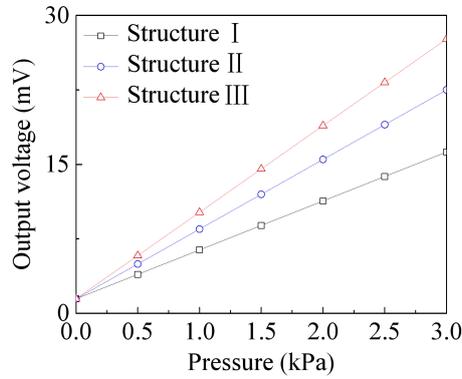

Figure 3. The input-output characteristic curves of pressure sensors with the three structure types.

### 3.1.2. *Effect of topline length on characteristics*

To analyze the effects of the topline length *d* on the characteristics of ITBM pressure sensors, the simulation models of ITBM pressure sensors with different *d* (including 280 μm, 300 μm, 320 μm and 340 μm) were established and their input-output characteristics were investigated as well. At $V_{DD}$= 5.0 V and in the pressure ranges of 0~3.0 kPa with a step 0.5 kPa, the input-output characteristic curves of the sensors are shown in Figure 4, where the output voltage increases with the decrease of *d*. In view of the simulation results, it can be seen that the sensor with *d*=280 μm achieves a maximum output voltagecomparing with that of the others, meaning a highest sensitivity in the size range of the study.

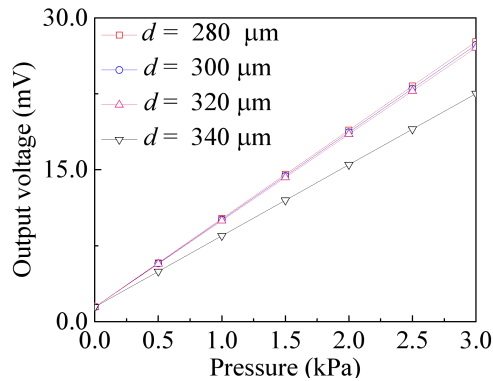

Figure 4. The input-output characteristic curves of proposed pressure sensors with different *d*.

### 3.2. *Fabrication Technology*

The proposed sensor is fabricated on a double-side polished silicon substrate with <100> orientation, as shown in Figure 5. The main processing steps of the micro-pressure sensor are as following: (a) cleaning a double-side polished silicon wafer with 4-inch n-type <100> orientation using a standard cleaning method; (b) growing a $SiO_2$ layer with a thickness of 40 nm by thermal oxidation; (c) first lithography to etch P$^+$ region window and injecting B-ions to form a heavily doped area as an ohmic contact; (d) second lithography to etch P$^-$ region window, thereafter injecting B-ions to form p-type diffused resistors and annealing at 900 ℃ for 30 min; (e) removing the SiO$_2$ layer and re-growing a SiO$_2$ layer of 500 nm, using third photolithography to form contact hole; (f) depositing a Al electrode layer, and then fourth photolithography to make a pattern electrode, after that, metalizing at 420 ℃ for 30 min to form an ohmic contact; (g) depositing a Si$_3$N$_4$ layer as a passivation layer, and fifth lithography to etch passivation layer window; (h) etching a depth of 7 μm on the top side of the chip by inductively coupled plasma (ICP) etching technology to form four trapezoidal beams; (i) seventh photolithography to etch the back side of the chip to form a silicon cup.

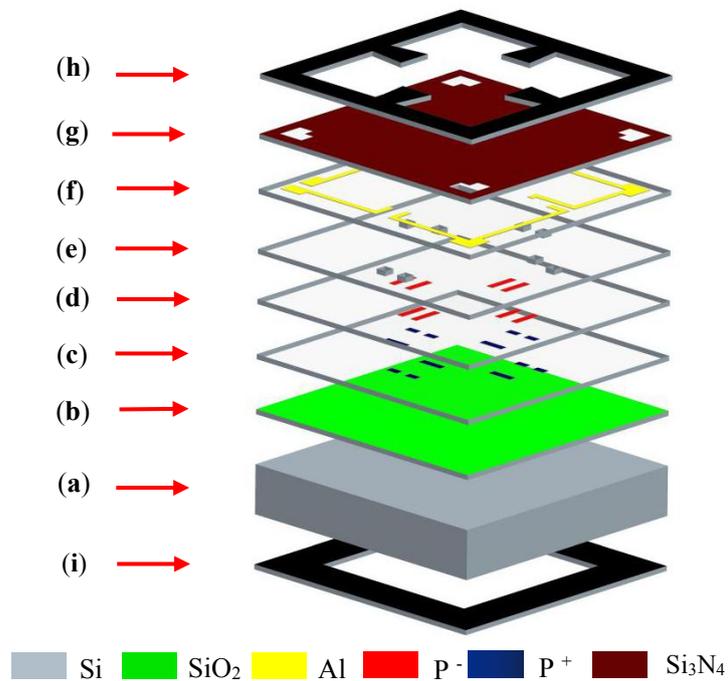

Figure 5. Main processing steps of the micro-pressure sensor: (a) cleaning Si wafer; (b) growing SiO$_2$ layer; (c) etching P$^+$ region window; (d) etching P$^-$ region window; (e) etching contact hole; (f) depositing Al electrode; (g) depositing passivation layer; (h) forming four trapezoidal beams; (i) fabricating silicon cup.

Figure 6 (a) and (b) show the front and back views of the chip, where the chip was bonded on a borosilicate glass to reduce the internal stress using anodic bonding technology, and then was packaged on a header by using gold wire to connect the Pad of chip. The photograph of packaged chip is shown in Figure 6 (c).

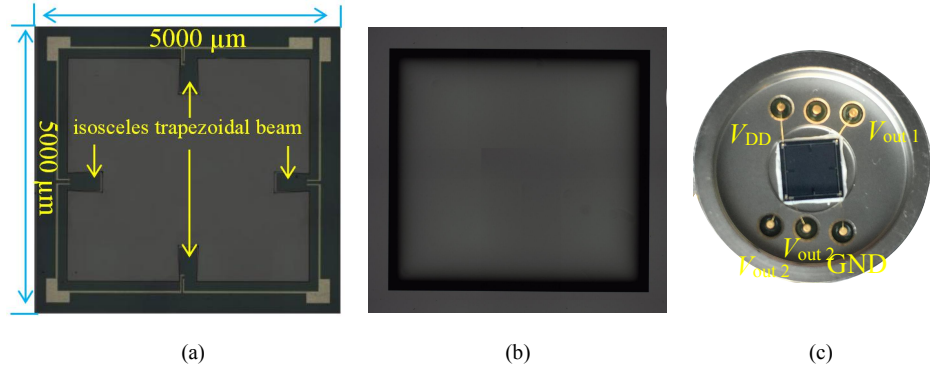

(a)                   (b)                   (c)

Figure 6. The photograph of chip: (a) the front view; (b) the back view; (c) the package.

## 4. Results and Discussion

To analyze static characteristics of fabricated pressure sensors, the relative tests were carried out by utilizing a pressure calibration system (Meson CPC 6000) in the pressure range of 0~3.0 kPa with a step 0.5 kPa, at room temperature and relative humidity (RH) of 43 %. When applying a voltage supply (RIGOL DP832A) to the sensors, the output voltages can be measured by using multimeter (Agilent 34401A). In addition, the temperature characteristics of the sensor are investigated by adopting a high-low temperature experiment chamber (GDJS-100LG-G) from -40℃ to 85℃.

### 4.1. *Effects of elastic silicon membrane and structural dimension on characteristics*

#### 4.1.1. *Effect of elastic silicon membrane on characteristics*

When $V_{DD}$ = 5.0 V, the relationship curves between the output voltages of the sensors with structures Ⅰ, Ⅱ and Ⅲ and the applied pressures were plotted in Figure 7, where the thicknesses of elastic elements for the sensors are 97 μm, 98 μm and 98 μm, and both of the beam thicknesses of the structures Ⅱ and Ⅲ are 7 μm, respectively. It can be seen that all of the output voltages for the sensors linearly increase with the pressures from 0 to 3.0 kPa. However, the ITBM pressure sensor presents a highest output voltage compared with that of the other structures at a pressure of 3.0 kPa, consisting with the previous simulation results. According to Eq.(7), the calculated sensitivities and linearity of the sensors with three structures are listed in Table 2. From the Table 2, it can be seen that the ITBM pressure sensor has a higher sensitivity and a lower linearity with respect to that of the pressure sensors with structures Ⅰ and Ⅱ, indicating that it is possible to balance the contradiction between the sensitivity and the linearity by using the pressure sensor with ITBM compared with the other type sensors.

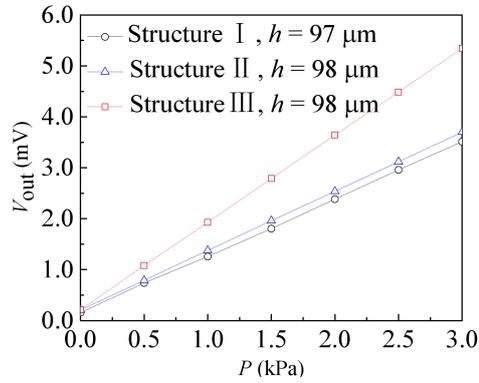

Figure 7. The input-output characteristic curves of the pressure sensors with different structures.

Table 2. The characteristics of the pressure sensors with different structures.

| Structure type | Sensitivity (mV/kPa) | linearity (% F.S.) |
|---|---|---|
| Structure Ⅰ | 1.12 | 0.77 |
| Structure Ⅱ | 1.16 | 0.14 |
| Structure Ⅲ | 1.71 | 0.12 |

#### 4.1.2. *Effect of topline length on characteristics*

Figure 8 shows the input-output characteristic curves of the proposed sensor at $V_{DD}$ = 5.0 V, corresponding to different $d$ of 280 μm, 300 μm, 320 μm and 340 μm, respectively. Due to the exhibiting of the deviation in fabrication process, the actual thicknesses of the elastic elements are 98 μm, 95 μm, 97 μm and 98 μm, respectively, and the thicknesses of corresponding beams are 7 μm. From Figure 8, it can be seen that the full scale output voltages of the ITBM pressure sensor increase as $d$ decreases. Moreover, the sensor achieves a minimum and a maximum sensitivity in the studied size ranges when $d$=340 μm and $d$=280 μm, respectively. It indicates that the simulation results relative to the sensors with different $d$ are in accord with that of the fabricated ITBM pressure sensors.

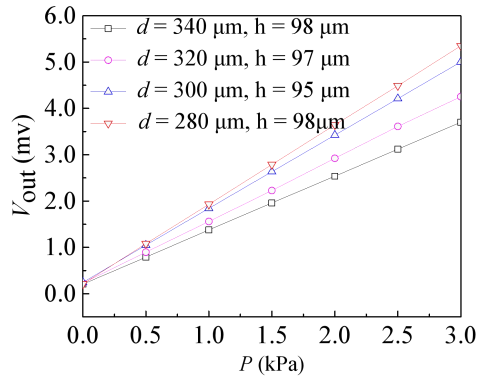

Figure 8. The input-output characteristic curves of the proposed pressure sensor with different $d$.

### 4.2. *The characteristics of ITBM pressure sensor*

Through thinning the thickness of the ITBM pressure sensor with $d$ = 280 μm, the final thicknesses of elastic element and beam are 42 μm and 7 μm, respectively. When exerting a supply voltage from 0 to 5.0 V with a step 1.0 V, the input-output characteristic curves of the ITBM pressure sensors were obtained as shown in Figure 9 (a). It can be seen that the full scale output voltage increases with supplied voltage $V_{DD}$. The sensitivities of the sensor under 1.0 V, 2.0 V, 3.0 V, 4.0 V and 5.0 V are 1.93mV/kPa, 3.85 mV/kPa, 5.79 mV/kPa, 7.75 mV/kP and 9.64 mV/kPa, respectively. When $V_{DD}$= 5.0 V, the sensor was repeatedly measured for three cycles in the pressure range of 0~3.0 kPa. Figure 9 (b) shows the input-output characteristic curves of the sensor, where the sensor has a sensitivity of 9.64 mV/kPa and realized a good linearity 0.09%F.S. as well as repeatability, with an average full scale output voltage of 28.92 mV. The relative parameters are shown in Table 3.

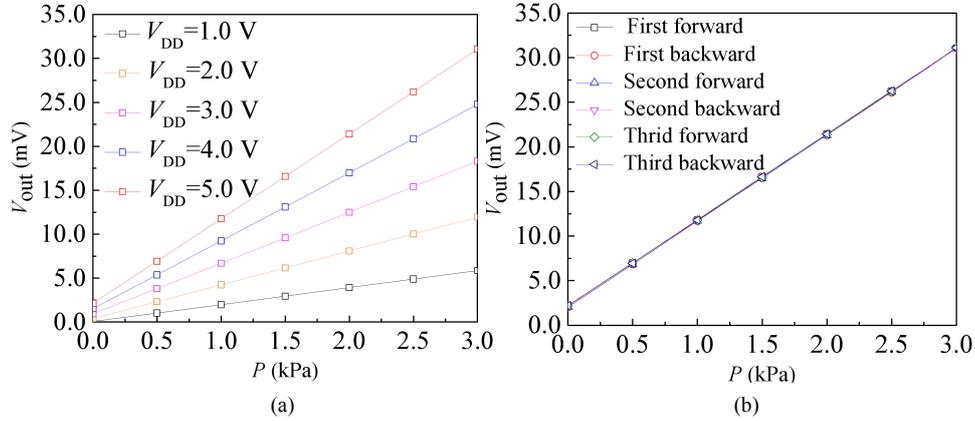

Figure 9. The input-output characteristic curves of the ITBM pressure sensor: (a) under different $V_{DD}$; (b) under $V_{DD}$ = 5.0 V.

Table 3. The characteristics of the ITBM pressure sensor under $V_{DD}$=5.0 V.

| Parameter | Value |
|---|---|
| Full scale output voltage (mV) | 28.92 |
| Sensitivity (mV/kPa) | 9.64 |
| linearity (%F.S.) | 0.09 |
| Hysteresis (%F.S.) | 0.15 |
| Repeatability (%F.S.) | 0.63 |
| Accuracy(% F.S.) | 0.67 |

When $V_{DD}$= 5.0 V, the effect of temperature on the sensitivity of the sensor was studied. Figure 10 shows the relationship curves between the sensitivity of the sensor and the temperature from -40℃ to 85℃, where the sensitivity decreases with the increase of temperature, presenting a negative temperature coefficient of ‐2180 ppm/℃.

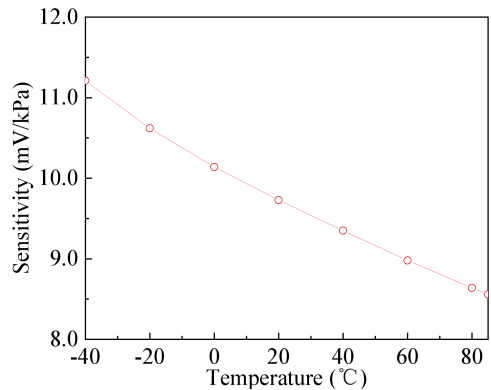

Figure 10. The relationship curves between the sensitivity of proposed sensor and temperature.

## 5. Conclusions

In summary, a pressure sensor with an ITBM structure was proposed in this study, where the ITBM structure consists of a square silicon membrane, four isosceles trapezoidal beams and four piezoresistors. To investigate the effects of the elastic silicon membrane on pressure sensitive characteristics, a simulation models based on ANSYS 15.0 software were installed. According to that, a micro-pressure sensor chips with different topline lengths were fabricated by micro-electro-mechanical system (MEMS) technology on a silicon wafer with <100> orientation. The test results show that comparing with the other structure sensors, the resulted sensor with the ITBM structure can achieve a better sensitivity of 9.64 mV/kPa and an excellent linearity of 0.09 %F.S in the range of 0~3.0 kPa at room temperature and a supply voltage of 5.0 V. In addition, it has a low pressure measurement less than 3.0 kPa and a super temperature coefficient of sensitivity (TCS) about - 2180 ppm/℃ from - 40 ℃ to 85 ℃. The study on the micro-pressure sensor with the ITBM structure provides a possibility to further improve the performance of micro-pressure sensors.

**6. Acknowledgments:** This work is supported by the National Natural Science Foundation of China (Grant No. 61971180, 61471159, 61006057).